\documentclass[11pt,reqno]{article}
\usepackage{jmlr2e}
\usepackage{fullpage,times,graphicx,amssymb,amsmath,amsfonts,bbm,psfrag,xcolor}

\usepackage{graphicx}
\usepackage{amssymb}
\usepackage{amsmath}
\usepackage{amsfonts}
\usepackage{bbm}
\usepackage{psfrag}
\usepackage{xcolor}
\usepackage{fullpage}
\usepackage{epstopdf}
\usepackage{graphicx}
\usepackage{times}
\usepackage{jmlr2e}
\usepackage{commath}
\usepackage{thmtools}
\usepackage{thm-restate}
\usepackage{times}
\usepackage{graphicx}
\usepackage{natbib}
\usepackage{dirtytalk}
\definecolor{ddarkbrown}{rgb}{0.5,0.2,0.05} \definecolor{bbluegray}{rgb}{0.05,0,0.5}

\usepackage[colorlinks,citecolor=bbluegray,linkcolor=ddarkbrown,urlcolor=blue,breaklinks]{hyperref}

\usepackage{subfig,float} 

\usepackage{mathtools}

\usepackage{algorithm,algcompatible,algpseudocode}
\usepackage{multicol,lipsum}
\algnewcommand{\Inputs}[1]{%
	\State \textbf{Inputs: \:}{#1}
}

\algnewcommand{\Output}[1]{%
	\State \textbf{Output: \:}{#1}
}
\algnewcommand{\Initialize}[1]{%
	\State \textbf{Initialize: \:}{#1}
}

\algnewcommand{\IIf}[1]{\State\algorithmicif\ #1\ \algorithmicthen}
\algnewcommand{\EndIIf}{\unskip\ \algorithmicend\ \algorithmicif}

\let \oldsection \section
\renewcommand{\section}{\vspace{3ex plus 1ex}\oldsection}

\newcommand{\BEAS}{\begin{eqnarray*}}
	\newcommand{\EEAS}{\end{eqnarray*}}
\newcommand{\BEA}{\begin{eqnarray}}
\newcommand{\EEA}{\end{eqnarray}}
\newcommand{\mb}{\mathbb}
\newcommand{\BEQ}{\begin{equation}}
\newcommand{\EEQ}{\end{equation}}
\newcommand{\BIT}{\begin{itemize}}
	\newcommand{\EIT}{\end{itemize}}
\newcommand{\BNUM}{\begin{enumerate}}
	\newcommand{\ENUM}{\end{enumerate}}

	\newcommand{\D}{\mathcal{D}}
	\newcommand{\Sk}{\mathcal{S}}
		\newcommand{\U}{\mathcal{U}}

	\newcommand{\R}{\mathbb{R}}

\newcommand{\BA}{\begin{array}}
	\newcommand{\EA}{\end{array}}

 \numberwithin{dummy}{section}

\numberwithin{mythm}{section}
\numberwithin{mydef}{section}
\numberwithin{myprop}{section}
\numberwithin{mylem}{section}
\numberwithin{mycor}{section}

\title{Accelerating Plug-and-Play Image Reconstruction \\via Multi-Stage Sketched Gradients}

\begin{document}
	\author{Junqi Tang \email jt814@cam.ac.uk\\
		\addr Department of Applied Mathematics and Theoretical Physics (DAMTP),\\ University of Cambridge
	}
	\editor{}
	
	
	\maketitle

\begin{abstract}
    In this work we propose a new paradigm for designing fast plug-and-play (PnP) algorithms using dimensionality reduction techniques. Unlike existing approaches which utilize stochastic gradient iterations for acceleration, we propose novel multi-stage sketched gradient iterations which first perform downsampling dimensionality reduction in the image space, and then efficiently approximate the true gradient using the sketched gradient in the low-dimensional space. This sketched gradient scheme can also be naturally combined with PnP-SGD methods for further improvement on computational complexity. As a generic acceleration scheme, it can be applied to accelerate any existing PnP/RED algorithm. Our numerical experiments on X-ray fan-beam CT demonstrate the remarkable effectiveness of our scheme, that a computational free-lunch can be obtained using this dimensionality reduction in the image space.
\end{abstract}

\section{Introduction}

Iterative model-based reconstruction algorithms have been the de-facto techniques for solving imaging inverse problems such as image deblurring/inpainting/superresolution and tomographic image reconstruction (for example X-ray CT, MRI and PET, etc). Due to their strengths in delivering data-consistent and robust reconstruction, these iterative solvers, especially when combined with advanced image priors \citep{dabov2007image, zhang2017beyond, tachella2020neural} in a \say{plug-and-play} (PnP) manner \citep{egiazarian2007compressed,venkatakrishnan2013plug, romano2017little, reehorst2018regularization}, can still thrive in current era where deep neural networks \citep{jin2017deep} have been successfully adopted in all these problems.

In this work, we propose a new paradigm of designing fast PnP algorithms utilizing the principle of dimension reduction in solving high-dimensional/resolution imaging inverse problems. Such imaging systems can be generally expressed as:
\begin{equation}\label{model}
   b = A x^\dagger + w,
\end{equation}
where $x^\dagger \in \mb{R}^d$, $A \in \mb{R}^{n \times d}$, $b \in \mb{R}^n$ and $w \in \mb{R}^n$ denote the ground truth image, the forward operator which models the measurement physics, the measurement data, and the measurement noise in the data (the noise can be data-dependent), respectively.

Traditionally, to obtain an estimate of the ground truth $x^\dagger$, we typically solve an optimization problem of the form:
\begin{equation}\label{1st-stage}
    x^\star \in \arg\min_{x \in \mb{R}^d} f(b, Ax) + g(x),
\end{equation}
where data fidelity term $f(b, Ax)$ is a convex function in $x$, and the most typical and widely used choice of the data fidelity would be the least-squares loss $\|b - Ax\|_2^2$. The $g(x)$ here is a regularization function which is usually convex to ensure provable convergence and robustness of the estimation, such as the TV regularization applied on image domain and $\ell_1$ regularization in wavelet/shearlet domain. This optimization problem can be efficiently solved by proximal splitting methods \citep{combettes2011proximal} for example FISTA \citep{beck2009fast} and ADMM \citep{boyd2011distributed}.

While these classical convex regularization approaches provide theoretically tractable solutions for inverse problems, they have been significantly outperformed by the PnP priors, constructed by advanced image denoisers or deep neural networks. The very first PnP algorithm (probably not very well-known) is actually proposed by \cite{egiazarian2007compressed}, which is a PnP stochastic approximation algorithm with BM3D as the denoiser. The PnP-ADMM of \cite{venkatakrishnan2013plug} and PnP-FISTA of \cite{kamilov2017plug} extend the classical methods ADMM and FISTA, replacing the proximal operator with the denoiser and have been widely applied in solving inverse problems since then. Then a very similar approach named regularization-by-denoising (RED) has been proposed by \cite{romano2017little,reehorst2018regularization}, which explicitly construct the regularization term using the denoiser and provide improved convenience in parameter tuning. Since a strong link between PnP and RED is established in \cite{cohen2021regularization} under the RED-PRO unifying framework, in this work we refer both plug-and-play and regularization-by-denoising as \say{PnP} for simplicity.

For large-scale problems, the PnP-ADMM and PnP-FISTA may require long computational time to obtain a good estimate. The PnP-SGD \citep{sun2019online} and stochastic PnP-ADMM methods \citep{tang2020fast,sun2020scalable} can provide significant acceleration compared to the deterministic PnP-ADMM/PnP-FISTA methods. In this work we propose a generic acceleration of PnP gradient methods using dimensionality-reduction/sketching in the image-space.

\section{Multi-Stage Sketched Plug-and-Play}

In this section we present our multi-stage sketched gradient framework PnP-MS2G. The sketching techniques have been widely applied in large-scale optimization especially the least-squares problems \citep{pilanci2017newton,2016_Pilanci_Iterative,2015_Pilanci_Randomized,tang2017exploiting, pmlr-v70-tang17a}. However, the author has found that such data-domain sketching methods are not efficient in imaging inverse problems, since very often the forward operator is relatively sparse, and even the most efficient sparse Johnson-Lindenstrauss transform \citep{woodruff2014sketching} cannot provide significant computational gain here since the sketched operator typically has similar sparsity as the full operator. If we use subsampling sketch which is the only practical data-domain sketch, the performance is similar or worse than SGD methods in imaging inverse problems.

Instead of using data-domain sketches, we propose to perform sketching in image-domain which appears to be much more effective and can be applied to generically accelerate PnP gradient methods such as PnP-FISTA and PnP-SGD.

\subsection{Algorithmic Framework}
Suppose the original objective reads:
\begin{equation}\label{obj_f}
    x^\star \in \arg\min_{x \in \mb{R}^d} f(b, Ax) + g(x),
\end{equation}
where $g(x)$ can be some implicit non-convex regularization constructed by the denoiser (we write this for the ease of presentation), then our sketched objective can be generally expressed as:
\begin{equation}\label{obj_s}
    x^\star \in \arg\min_{x \in \mb{R}^d} f(b, A_s \Sk(x)) + g(x),
\end{equation}
where $\Sk(\cdot): \R^d \rightarrow \R^m$ ($m < d$) being the sketching/downsampling operator, while $A_s \in \R^{n \times m}$ is the forward operator discretized on the reduced image space. We found that such a scheme provides a remarkably efficient approximation of the solution. We present our PnP-MS2G framework in Algorithm 1, where we denote $\D$ as the denoiser, $\Sk$ as the sketching operator, and $\U$ as the upsampling operator.

\begin{algorithm}[t]\label{AA}
   \caption{--- Plug-and-Play with Multi-Stage Sketched Gradients (PnP-MS2G)}
\begin{algorithmic}

   \State  {\bfseries Initialization:} $x_0 = y_0 \in \R^d$, number of stages $K$, sketch-size $[m_1,...,m_K]$, sketched forward operator $[A_{s_1},... A_{s_k}]$, sketching operators $[\Sk_1,...,\Sk_K]$, up-sampling operators $[\U_1,...,\U_K]$, number of inner-loops for each stage $[N_1,...,N_K]$, step-size sequence $[\eta_1,...,\eta_{\sum_{k=1}^K N_k}]$, momentum sequence $[a_1,...,a_{\sum_{k=1}^K N_k}]$, $\alpha \in (0, 1]$,  iteration counter $i = 0$
   \For{$k = 1$ {\bfseries to} $K$}
   \For{$j =1$ {\bfseries to} $N_k$}
\State $i \leftarrow i + 1$
 \State Compute the sketch:
   \State $v = \Sk_k(y_i)$
   \State Compute gradient estimate $G$:
    \State {\color{blue}Option 1:}  $G := \triangledown f_{v}(b, A_{s_k} v)$; \ \ \   or a stochastic gradient estimation of it:
    \State {\color{blue}Option 2:} $G := \triangledown f_{v}(M_ib, M_i A_{s_k} v)$ where $M_i$ is a randomly sampled minibatch of $I_{n \times n}$.
   \State        $z_{i+1} =  y_i - \eta_i \U_k G,$
   \State     $x_{i+1} =  (1-\alpha)z_{i+1} + \alpha \D(z_{i+1}),$

    \State $y_{i+1} = x_{i+1} + a_i (x_{i+1} - x_i)$
  
   \EndFor
   \EndFor
\State Output $x_{i+1}$
\end{algorithmic}
\end{algorithm} 

To explain the motivation and derivation of Algorithm 1, we start by illustrating here a concrete example where the data-fidelity is the least-squares. Noting that the PnP proximal gradient descent iteration can be written as:
\begin{equation}\label{pnp}
     x_{k+1} =  \D[x_k - \eta \cdot (A^TA x_k - A^T b)],
\end{equation}
where $\D(\cdot)$ denotes the denoiser, which can be a denoising algorithm such as NLM/BM3D/TNRD, or a classical proximal operator of some convex regularization (such as TV-prox), or a pretrained denoising deep network such as (DnCNN). Then our sketched gradient can be written as:
\begin{equation}
    x_{k+1} =  \D[x_k - \eta \cdot \U(A_s^TA_s \Sk(x_k) - A_s^Tb)],
\end{equation}
where $\U(\cdot)$ denotes the upsamling operator. Numerically we found that off-the-shelf up/down-sampling algorithms such as the bi-cubic interpolation suffice to provide us good estimates of the true gradients. Using this scheme, an efficient approximation of the true gradient can be obtained since $A_s$ only takes a fraction of the computation of $A$, and usually $\U$ and $\Sk$ can be very efficiently computed. 

To further reduce the computational complexity, we can also utilize stochastic gradient estimate:
\begin{equation}
    x_{k+1} =  \D[y_k - \eta_k \cdot \U((M_kA_s)^T M_kA_s \Sk(y_k) - (M_kA_s)^T M_kb)]
\end{equation}
where $M_k$ is uniformly sampled minibatch of $I_{n \times n}$ here for computing the stochastic gradient. Here we use a vanilla minibatch stochastic gradient estimator. We can also choose here those advanced stochastic variance-reduced gradient estimator \citep{2012_Roux_Stochastic,johnson2013accelerating,defazio2014saga,driggs2020spring} for potentially even faster convergence. Note that the stochastic gradient iterations do not necessary provide improved performance in all inverse problems -- there are cases of inverse problems for which the stochastic methods do not provide faster convergence compared to deterministic methods \citep{tang2020practicality}. For compressed sensing MRI where we use subsampled FFT as forward operator, and even in CT/PET reconstruction if we use Non-uniform FFT \citep{fessler2003nonuniform,tang2016non} for fast approximation of the Radon transform, the stochastic gradient is not applicable since it will break the fast computational structure of FFT/NUFFT.

In Algorithm 1 we present our PnP-MS2G framework, where we typically start by an aggressive sketch $\{A_{s_1}, \Sk_{1}\}$ with $m_1 \ll d$ for very fast initial convergence, and then for later stages we switch to medium size sketches $\{A_{s_k}, \Sk_{k}\}$ with $m_k < d$ which are increasingly more conservative, to reach a similar reconstruction accuracy as the unsketched counterpart. Our algorithmic framework admit both deterministic gradient (Option 1) and stochastic gradient (Option 2), with momentum acceleration for which we recommend FISTA-type momentum \citep{beck2009fast,chambolle2015convergence}. For the denoising step we integrate the unifying RED-PRO \citep{cohen2021regularization} framework here which includes both PnP (if we choose $\alpha = 1$) and RED as special cases.
 \begin{figure}[t]
   \centering

    {\includegraphics[width= .99\textwidth]{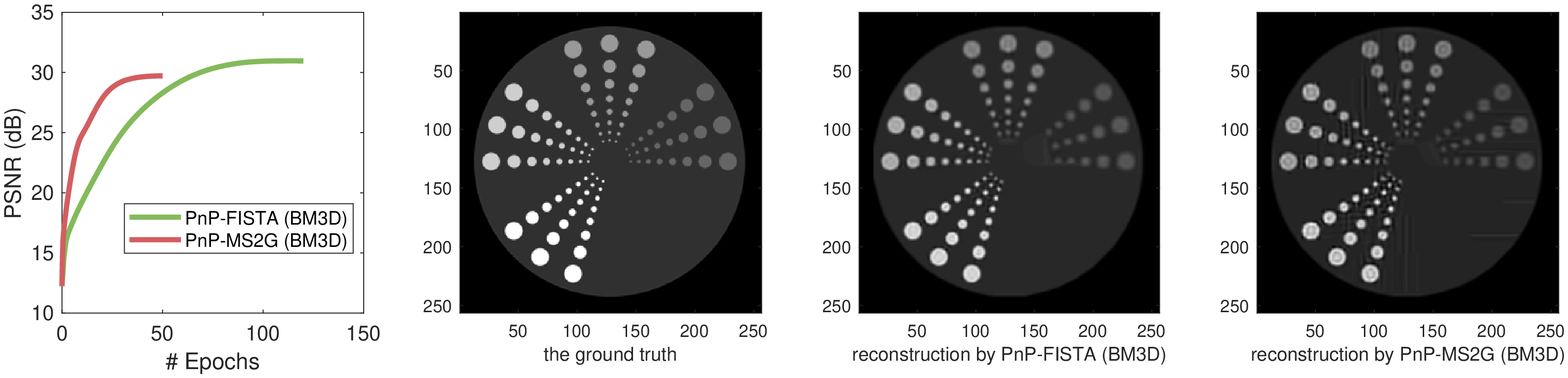}}\\
    {\includegraphics[width= .99\textwidth]{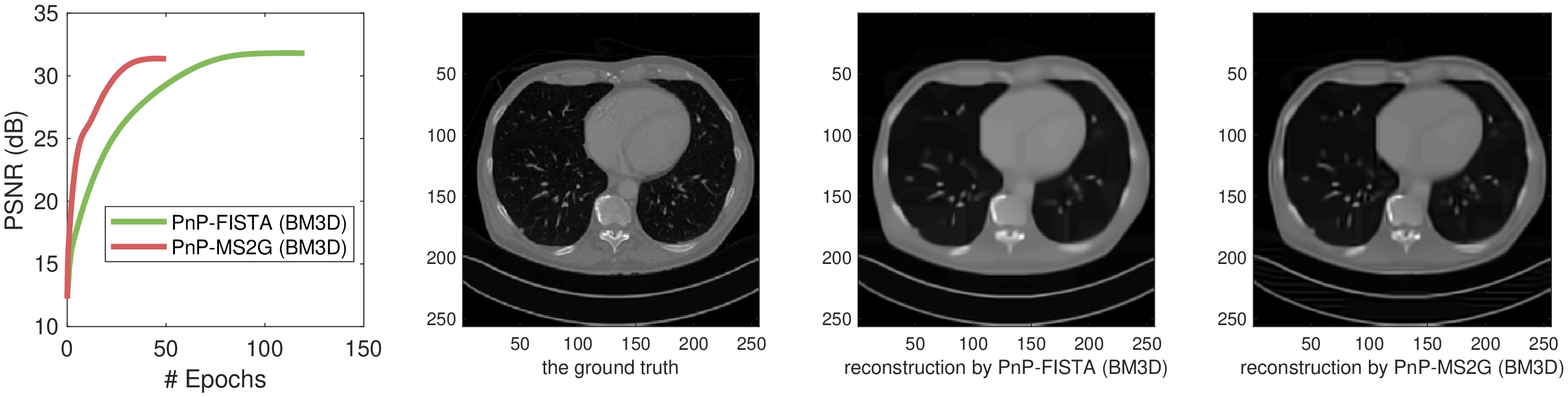}}\\
    {\includegraphics[width= .99\textwidth]{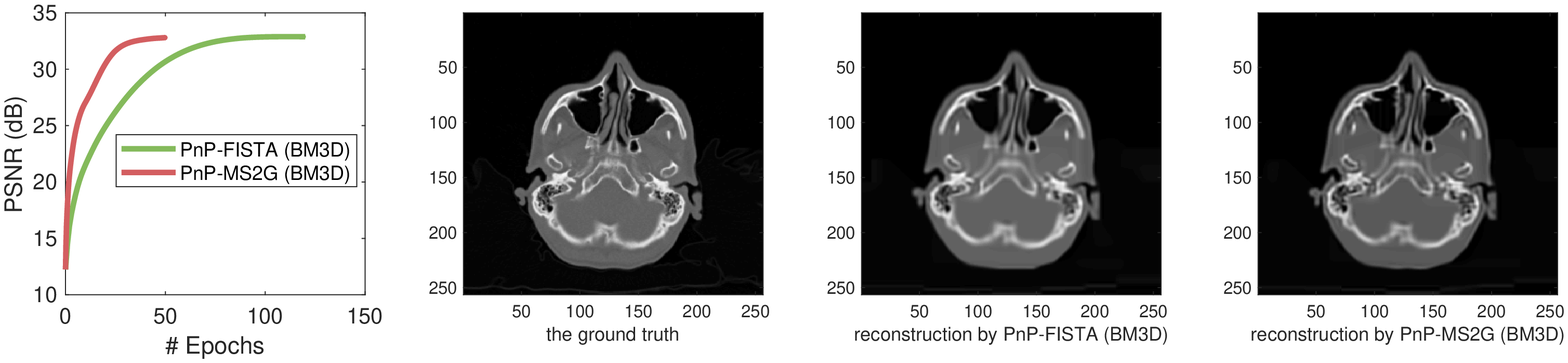}}

   \caption{Numerical results for {\color{purple}Sparse-View} CT, with $I_0 = 1 \times 10^{5.5}$, $A \in \R^{20520 \times 65536}$, 90 view equally space from $[0, 360)$ degrees}
   \label{f1}
\end{figure}

We also wish to point out that in our sketching framework both the denoiser $\D$, the upsampling function $\U$ and sketching function $\Sk$ can be parameterized as deep (convolutional) neural-networks and trained either in a recursive or end-to-end manner, resulting in a new efficient deep-unrolling scheme \citep{adler2018learned,tang2021stochastic}.

In the multi-sketch framework presented here we gradually increase the sketch size $m$ throughout stages. A seemly plausible alternative could be modifying the iterative-Hessian-sketch framework proposed by \cite{2016_Pilanci_Iterative} to sketch on the image domain. However in our initial experiments we found such a scheme is not as efficient as our PnP-MS2G and hence we do not report this alternative scheme here.

\section{Numerical Experiments}

In this section we present some numerical results for the proof-of-concept. We consider 3 fan-beam X-ray CT examples, sparse-view CT, low-dose CT, and limited angle CT. We simulate the noisy measurements (with Poisson noise):
\begin{equation}
    b \sim \mathrm{Poisson}(I_0 e^{-Ax^\dagger}),
\end{equation}
and take the logarithmic of the data. All the experiments here are implemented and executed using MATLAB R2018b.

 \begin{figure}[t]
   \centering

    {\includegraphics[width= .99\textwidth]{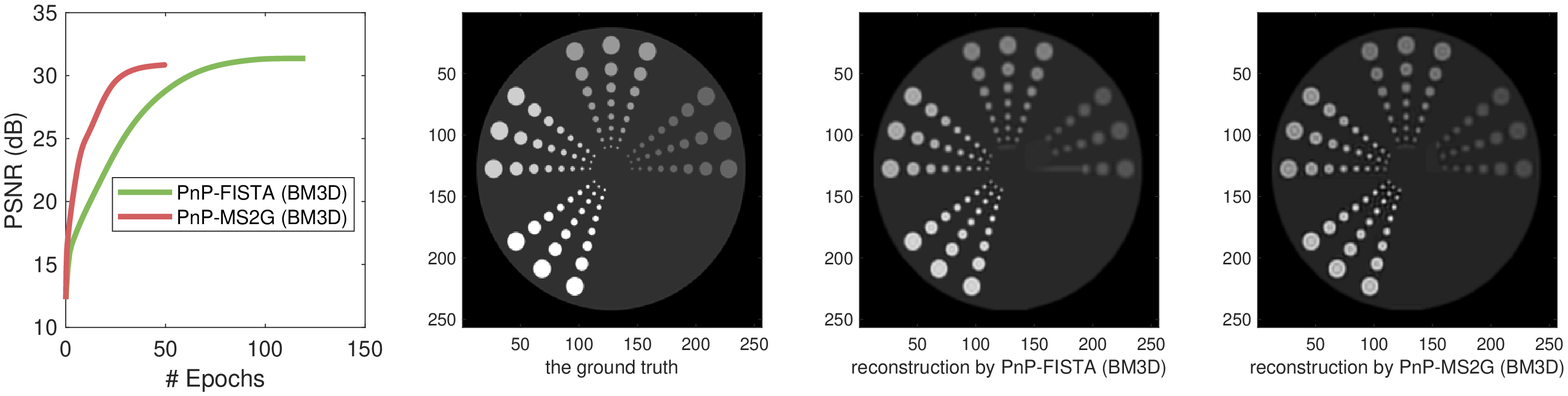}}\\
    {\includegraphics[width= .99\textwidth]{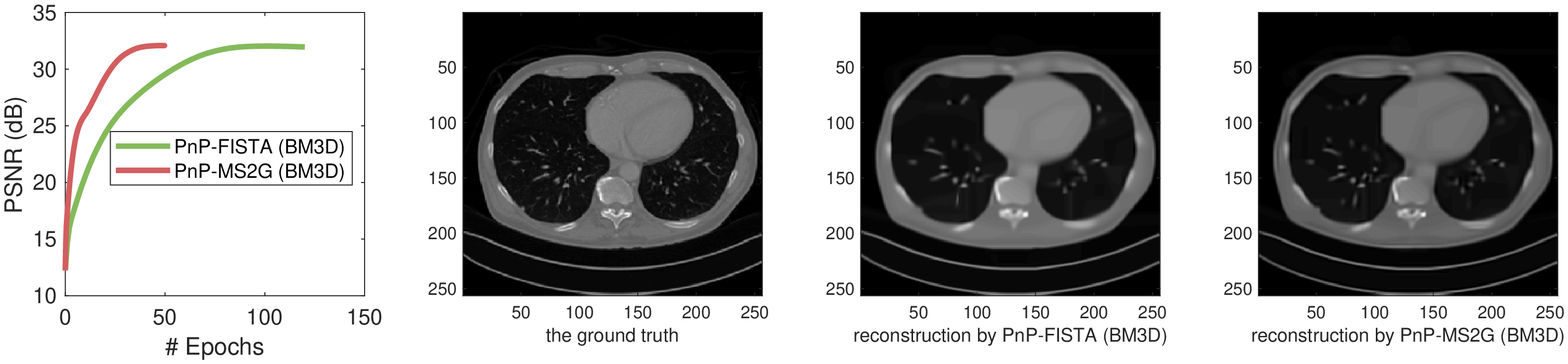}}\\
    {\includegraphics[width= .99\textwidth]{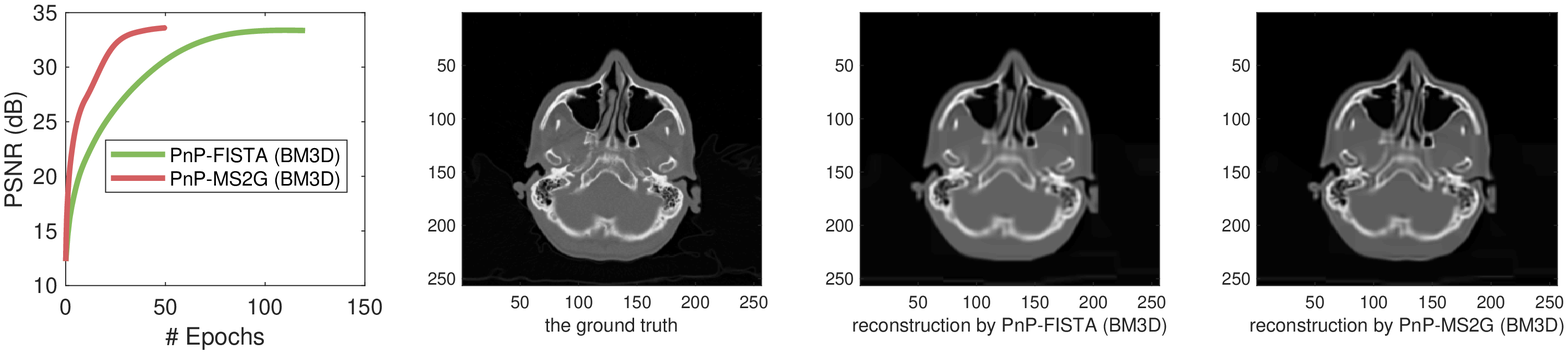}}

   \caption{Numerical results for {\color{purple}Low-Dose} CT, with $I_0 = 1 \times 10^{3.5}$, $A \in \R^{82080 \times 65536}$, 360 view equally space from $[0, 360)$ degrees}
   \label{f2}
\end{figure}

 \begin{figure}[t]
   \centering

    {\includegraphics[width= .99\textwidth]{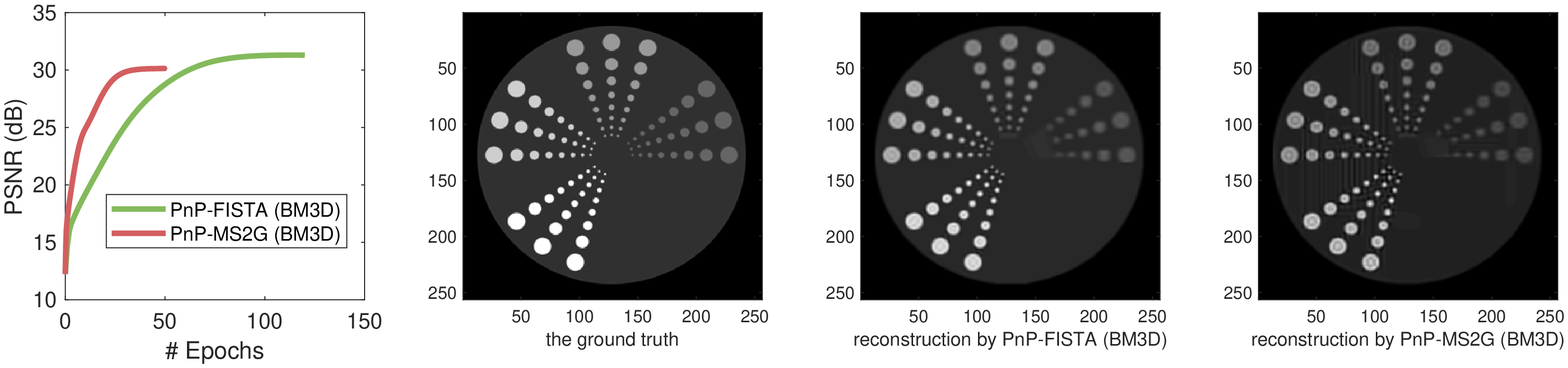}}\\
    {\includegraphics[width= .99\textwidth]{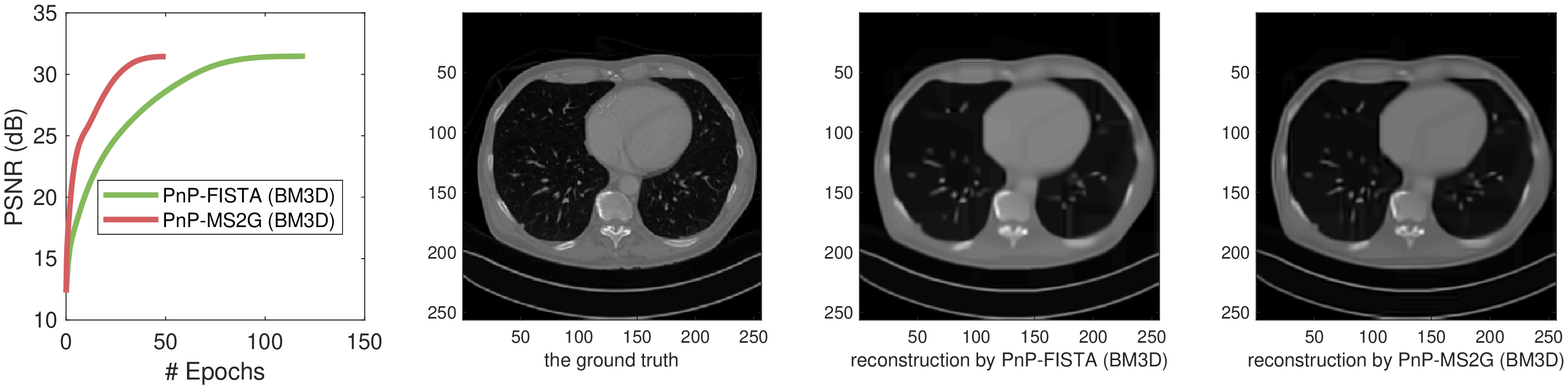}}\\
    {\includegraphics[width= .99\textwidth]{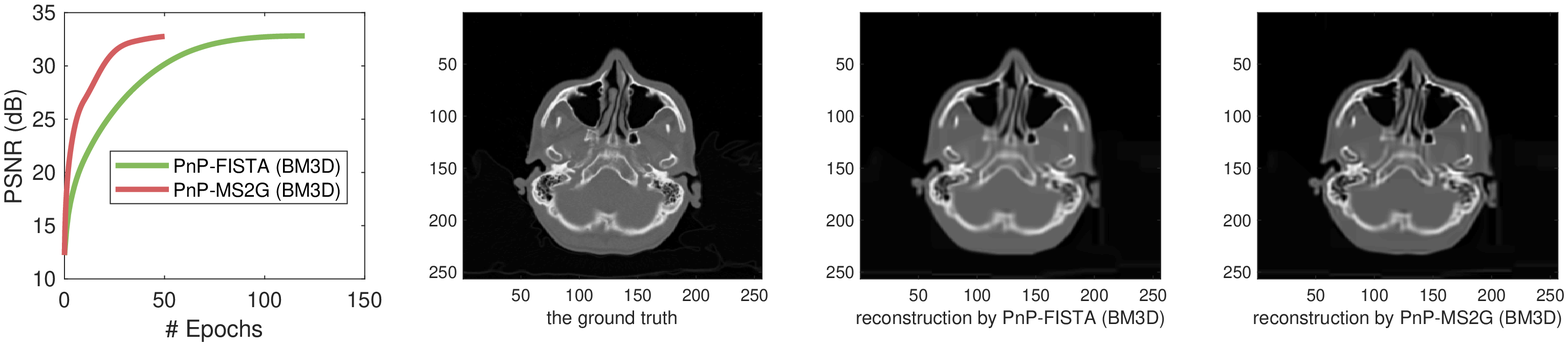}}

 \caption{Numerical results for {\color{purple}Limited-Angle} CT, with $I_0 = 1 \times 10^{5.5}$, $A \in \R^{20520 \times 65536}$, 90 view equally space from $[0, 180)$ degrees}
   \label{f3}
\end{figure}

In all the experiments we use the BM3D \citep{dabov2007image} as the denoiser for the PnP methods. We test the Option 1 (deterministic gradient) of PnP-MS2G and compare it to PnP-FISTA \citep{kamilov2017plug}.  For the up-sampling operator $\U$ and sketching operator $\Sk$, we using the MATLAB \texttt{imresize} function with \texttt{bicubic} interpolation. We use 3 testing images (both sized $256 \times 256$): a phantom image with disks of various sizes, a head image, and a chest image. We use a fixed step-size $\eta = O(\frac{1}{L})$ where $L$ is the gradient Lipschitz constant of the data-fidelity function and the momentum parameter $a_i = \frac{i-1}{i+3}$ adopted from the \say{convergent-FISTA} algorithm \citep{chambolle2015convergence} which has provable convergence on the variable. For simplicity we choose $\alpha = 1$ for PnP-MS2G. Meanwhile for PnP-MS2G we choose the number of stages as 2, where for the first stage we use 16-fold reduction on the dimension ($m_1 = \frac{d}{16}$), and for the second stage we use 4-fold reduction ($m_2 = \frac{d}{4}$)

We present our sparse view CT results in Figure \ref{f1}, low-dose CT results in Figure \ref{f2}, and limited angle CT results in Figure \ref{f3}. We can observe that for all these examples our PnP-MS2G with deterministic gradient (Option 1) achieves almost the same reconstruction accuracy in PSNR compared to its full counterpart PnP-FISTA, while only requires a fraction of the computation of it. These results demonstrates that our proposed scheme for reducing the computational cost of PnP gradient methods is remarkably effective.

\section{Conclusion}

In this work we proposed a generic acceleration framework for accelerating gradient-based PnP/RED methods, exploiting the low-dimensional structure of the images which allowed us to efficiently approximate the gradient in the high-dimensional image space via dimensionality reduction schemes. In principle, our proposed multi-stage sketching framework can be readily applied to accelerate any existing determinsitic/stochastic PnP/RED/Unrolling scheme. Our numerical experiments on sparse-view/low-dose/limited-angle CT demonstrate the effectiveness and huge potential of our framework.

This generic acceleration scheme can be easily applied to the cases where the non-uniform FFT (NUFFT) is used to provide fast computation of the forward operator (the NUFFT acceleration can be applied to both MRI/CT/PET). In such cases the stochastic gradients cannot be efficiently applied since it will break the fast computation of FFT.

\bibliography{main.bib}

\begin{thebibliography}{32}
\providecommand{\natexlab}[1]{#1}
\providecommand{\url}[1]{\texttt{#1}}
\expandafter\ifx\csname urlstyle\endcsname\relax
  \providecommand{\doi}[1]{doi: #1}\else
  \providecommand{\doi}{doi: \begingroup \urlstyle{rm}\Url}\fi

\bibitem[Adler and {\"O}ktem(2018)]{adler2018learned}
Jonas Adler and Ozan {\"O}ktem.
\newblock Learned primal-dual reconstruction.
\newblock \emph{IEEE transactions on medical imaging}, 37\penalty0
  (6):\penalty0 1322--1332, 2018.

\bibitem[Beck and Teboulle(2009)]{beck2009fast}
A.~Beck and M.~Teboulle.
\newblock Fast gradient-based algorithms for constrained total variation image
  denoising and deblurring problems.
\newblock \emph{IEEE Transactions on Image Processing}, 18\penalty0
  (11):\penalty0 2419--2434, 2009.

\bibitem[Boyd et~al.(2011)Boyd, Parikh, Chu, Peleato, Eckstein,
  et~al.]{boyd2011distributed}
Stephen Boyd, Neal Parikh, Eric Chu, Borja Peleato, Jonathan Eckstein, et~al.
\newblock Distributed optimization and statistical learning via the alternating
  direction method of multipliers.
\newblock \emph{Foundations and Trends{\textregistered} in Machine learning},
  3\penalty0 (1):\penalty0 1--122, 2011.

\bibitem[Chambolle and Dossal(2015)]{chambolle2015convergence}
Antonin Chambolle and Ch~Dossal.
\newblock On the convergence of the iterates of the “fast iterative
  shrinkage/thresholding algorithm”.
\newblock \emph{Journal of Optimization theory and Applications}, 166\penalty0
  (3):\penalty0 968--982, 2015.

\bibitem[Cohen et~al.(2021)Cohen, Elad, and Milanfar]{cohen2021regularization}
Regev Cohen, Michael Elad, and Peyman Milanfar.
\newblock Regularization by denoising via fixed-point projection (red-pro).
\newblock \emph{SIAM Journal on Imaging Sciences}, 14\penalty0 (3):\penalty0
  1374--1406, 2021.

\bibitem[Combettes and Pesquet(2011)]{combettes2011proximal}
Patrick~L Combettes and Jean-Christophe Pesquet.
\newblock Proximal splitting methods in signal processing.
\newblock In \emph{Fixed-point algorithms for inverse problems in science and
  engineering}, pages 185--212. Springer, 2011.

\bibitem[Dabov et~al.(2007)Dabov, Foi, Katkovnik, and
  Egiazarian]{dabov2007image}
K~Dabov, A~Foi, V~Katkovnik, and K~Egiazarian.
\newblock Image denoising by sparse 3-d transform-domain collaborative
  filtering.
\newblock \emph{IEEE transactions on image processing: a publication of the
  IEEE Signal Processing Society}, 16\penalty0 (8):\penalty0 2080--2095, 2007.

\bibitem[Defazio et~al.(2014)Defazio, Bach, and
  Lacoste-Julien]{defazio2014saga}
A.~Defazio, F.~Bach, and S.~Lacoste-Julien.
\newblock Saga: A fast incremental gradient method with support for
  non-strongly convex composite objectives.
\newblock In \emph{Advances in Neural Information Processing Systems}, pages
  1646--1654, 2014.

\bibitem[Driggs et~al.(2021)Driggs, Tang, Liang, Davies, and
  Schonlieb]{driggs2020spring}
Derek Driggs, Junqi Tang, Jingwei Liang, Mike Davies, and Carola-Bibiane
  Schonlieb.
\newblock A stochastic proximal alternating minimization for nonsmooth and
  nonconvex optimization.
\newblock \emph{SIAM Journal on Imaging Sciences}, 14\penalty0 (4):\penalty0
  1932--1970, 2021.

\bibitem[Egiazarian et~al.(2007)Egiazarian, Foi, and
  Katkovnik]{egiazarian2007compressed}
Karen Egiazarian, Alessandro Foi, and Vladimir Katkovnik.
\newblock Compressed sensing image reconstruction via recursive spatially
  adaptive filtering.
\newblock In \emph{2007 IEEE International Conference on Image Processing},
  volume~1, pages I--549. IEEE, 2007.

\bibitem[Fessler and Sutton(2003)]{fessler2003nonuniform}
Jeffrey~A Fessler and Bradley~P Sutton.
\newblock Nonuniform fast fourier transforms using min-max interpolation.
\newblock \emph{IEEE transactions on signal processing}, 51\penalty0
  (2):\penalty0 560--574, 2003.

\bibitem[Jin et~al.(2017)Jin, McCann, Froustey, and Unser]{jin2017deep}
Kyong~Hwan Jin, Michael~T McCann, Emmanuel Froustey, and Michael Unser.
\newblock Deep convolutional neural network for inverse problems in imaging.
\newblock \emph{IEEE Transactions on Image Processing}, 26\penalty0
  (9):\penalty0 4509--4522, 2017.

\bibitem[Johnson and Zhang(2013)]{johnson2013accelerating}
Rie Johnson and Tong Zhang.
\newblock Accelerating stochastic gradient descent using predictive variance
  reduction.
\newblock In \emph{Advances in neural information processing systems}, pages
  315--323, 2013.

\bibitem[Kamilov et~al.(2017)Kamilov, Mansour, and Wohlberg]{kamilov2017plug}
Ulugbek~S Kamilov, Hassan Mansour, and Brendt Wohlberg.
\newblock A plug-and-play priors approach for solving nonlinear imaging inverse
  problems.
\newblock \emph{IEEE Signal Processing Letters}, 24\penalty0 (12):\penalty0
  1872--1876, 2017.

\bibitem[Pilanci and Wainwright(2015)]{2015_Pilanci_Randomized}
M.~Pilanci and M.~J. Wainwright.
\newblock Randomized sketches of convex programs with sharp guarantees.
\newblock \emph{Information Theory, IEEE Transactions on}, 61\penalty0
  (9):\penalty0 5096--5115, 2015.

\bibitem[Pilanci and Wainwright(2016)]{2016_Pilanci_Iterative}
M.~Pilanci and M.~J. Wainwright.
\newblock Iterative hessian sketch: Fast and accurate solution approximation
  for constrained least-squares.
\newblock \emph{Journal of Machine Learning Research}, 17\penalty0
  (53):\penalty0 1--38, 2016.

\bibitem[Pilanci and Wainwright(2017)]{pilanci2017newton}
Mert Pilanci and Martin~J Wainwright.
\newblock Newton sketch: A near linear-time optimization algorithm with
  linear-quadratic convergence.
\newblock \emph{SIAM Journal on Optimization}, 27\penalty0 (1):\penalty0
  205--245, 2017.

\bibitem[Reehorst and Schniter(2018)]{reehorst2018regularization}
Edward~T Reehorst and Philip Schniter.
\newblock Regularization by denoising: Clarifications and new interpretations.
\newblock \emph{IEEE Transactions on Computational Imaging}, 5\penalty0
  (1):\penalty0 52--67, 2018.

\bibitem[Romano et~al.(2017)Romano, Elad, and Milanfar]{romano2017little}
Yaniv Romano, Michael Elad, and Peyman Milanfar.
\newblock The little engine that could: Regularization by denoising (red).
\newblock \emph{SIAM Journal on Imaging Sciences}, 10\penalty0 (4):\penalty0
  1804--1844, 2017.

\bibitem[Roux et~al.(2012)Roux, Schmidt, and Bach]{2012_Roux_Stochastic}
Nicolas~L. Roux, Mark Schmidt, and Francis~R. Bach.
\newblock A stochastic gradient method with an exponential convergence \_rate
  for finite training sets.
\newblock In F.~Pereira, C.~J.~C. Burges, L.~Bottou, and K.~Q. Weinberger,
  editors, \emph{Advances in Neural Information Processing Systems 25}, pages
  2663--2671. Curran Associates, Inc., 2012.

\bibitem[Sun et~al.(2019)Sun, Wohlberg, and Kamilov]{sun2019online}
Yu~Sun, Brendt Wohlberg, and Ulugbek~S Kamilov.
\newblock An online plug-and-play algorithm for regularized image
  reconstruction.
\newblock \emph{IEEE Transactions on Computational Imaging}, 2019.

\bibitem[Sun et~al.(2020)Sun, Wu, Wohlberg, and Kamilov]{sun2020scalable}
Yu~Sun, Zihui Wu, Brendt Wohlberg, and Ulugbek~S Kamilov.
\newblock Scalable plug-and-play admm with convergence guarantees.
\newblock \emph{arXiv preprint arXiv:2006.03224}, 2020.

\bibitem[Tachella et~al.(2021)Tachella, Tang, and Davies]{tachella2020neural}
Juli{\'a}n Tachella, Junqi Tang, and Mike Davies.
\newblock The neural tangent link between cnn denoisers and non-local filters.
\newblock \emph{IEEE/CVF Conference on Computer Vision and Pattern
  Recognition}, 2021.

\bibitem[Tang(2016)]{tang2016non}
Junqi Tang.
\newblock The non-uniform fast fourier transform in computed tomography.
\newblock \emph{arXiv preprint arXiv:1605.05231}, 2016.

\bibitem[Tang and Davies(2020)]{tang2020fast}
Junqi Tang and Mike Davies.
\newblock A fast stochastic plug-and-play admm for imaging inverse problems.
\newblock \emph{arXiv preprint arXiv:2006.11630}, 2020.

\bibitem[Tang et~al.(2017{\natexlab{a}})Tang, Golbabaee, and
  Davies]{tang2017exploiting}
Junqi Tang, Mohammad Golbabaee, and Mike Davies.
\newblock Exploiting the structure via sketched gradient algorithms.
\newblock In \emph{2017 IEEE Global Conference on Signal and Information
  Processing (GlobalSIP)}, pages 1305--1309. IEEE, 2017{\natexlab{a}}.

\bibitem[Tang et~al.(2017{\natexlab{b}})Tang, Golbabaee, and
  Davies]{pmlr-v70-tang17a}
Junqi Tang, Mohammad Golbabaee, and Mike~E. Davies.
\newblock Gradient projection iterative sketch for large-scale constrained
  least-squares.
\newblock In \emph{Proceedings of the 34th International Conference on Machine
  Learning}, volume~70 of \emph{Proceedings of Machine Learning Research},
  pages 3377--3386. PMLR, 2017{\natexlab{b}}.

\bibitem[Tang et~al.(2020)Tang, Egiazarian, Golbabaee, and
  Davies]{tang2020practicality}
Junqi Tang, Karen Egiazarian, Mohammad Golbabaee, and Mike Davies.
\newblock The practicality of stochastic optimization in imaging inverse
  problems.
\newblock \emph{IEEE Transactions on Computational Imaging}, 6:\penalty0
  1471--1485, 2020.

\bibitem[Tang et~al.(2021)Tang, Mukherjee, and Schonlieb]{tang2021stochastic}
Junqi Tang, Subhadip Mukherjee, and Carola-Bibiane Schonlieb.
\newblock Stochastic primal-dual deep unrolling for imaging inverse problems.
\newblock \emph{arXiv preprint arXiv:2110.10093}, 2021.

\bibitem[Venkatakrishnan et~al.(2013)Venkatakrishnan, Bouman, and
  Wohlberg]{venkatakrishnan2013plug}
Singanallur~V Venkatakrishnan, Charles~A Bouman, and Brendt Wohlberg.
\newblock Plug-and-play priors for model based reconstruction.
\newblock In \emph{2013 IEEE Global Conference on Signal and Information
  Processing}, pages 945--948. IEEE, 2013.

\bibitem[Woodruff et~al.(2014)]{woodruff2014sketching}
David~P Woodruff et~al.
\newblock Sketching as a tool for numerical linear algebra.
\newblock \emph{Foundations and Trends{\textregistered} in Theoretical Computer
  Science}, 10\penalty0 (1--2):\penalty0 1--157, 2014.

\bibitem[Zhang et~al.(2017)Zhang, Zuo, Chen, Meng, and Zhang]{zhang2017beyond}
Kai Zhang, Wangmeng Zuo, Yunjin Chen, Deyu Meng, and Lei Zhang.
\newblock Beyond a gaussian denoiser: Residual learning of deep cnn for image
  denoising.
\newblock \emph{IEEE Transactions on Image Processing}, 26\penalty0
  (7):\penalty0 3142--3155, 2017.

\end{thebibliography}

\end{document}